\documentclass[letterpaper,10 pt,conference]{ieeeconf}  
\makeatletter
\let\NAT@parse\undefined
\makeatother

\IEEEoverridecommandlockouts                              
\usepackage{graphicx}
\usepackage{subfigure}
\usepackage{booktabs} % for professional tables
\usepackage{float}
\usepackage{hyperref}
\usepackage{algorithm}
\usepackage{algpseudocodex}

\usepackage[utf8]{inputenc} % allow utf-8 input
\usepackage{amsmath}
\usepackage{amssymb}
\usepackage{mathtools}
\usepackage{bm}
\usepackage{nicefrac}    
\usepackage{xcolor}         % colors
\usepackage{balance}

\usepackage{colorlib}
\colorlet{link}{text1}
\hypersetup{
    colorlinks=true,
    linkcolor=link,
    filecolor=link,      
    urlcolor=link,
    citecolor=link,
}

\let\citet\cite
\newcommand{\setup}{MAMQL}

\title{Multi-Agent Inverse Q-Learning from Demonstrations}

\author{%
    Nathaniel Haynam$^{1}$ \quad
    Adam Khoja$^{1}$ \quad 
    Dhruv Kumar$^{1}$ \quad
    \\
    Vivek Myers$^{1}$ \quad
    Erdem B{\i}y{\i}k$^{1,2}$
    \thanks{
        $^{1}$Electrical Engineering and Computer Sciences, UC Berkeley
    }
    \thanks{
        $^{2}$Computer Science, University of Southern California
    }
}

\usepackage[capitalize]{cleveref}
\usepackage{pgfplots}
\usepackage{pgfplotstable}
\usepgfplotslibrary{fillbetween}

\begin{document}

\maketitle

\newcommand{\argmin}{\operatornamewithlimits{arg\,min}}
\newcommand{\argmax}{\operatornamewithlimits{arg\,max}}
\DeclarePairedDelimiter\abs{\lvert}{\rvert}
\DeclarePairedDelimiter\norm{\lVert}{\rVert}
\DeclarePairedDelimiter\dotp{\langle}{\rangle}

\begin{abstract}
	When reward functions are hand-designed, deep reinforcement learning algorithms often suffer from reward misspecification, causing them to learn suboptimal policies in terms of the intended task objectives. In the single-agent case, inverse reinforcement learning (IRL) techniques attempt to address this issue by inferring the reward function from expert demonstrations. However, in multi-agent problems, misalignment between the learned and true objectives is exacerbated due to increased environment non-stationarity and variance that scales with multiple agents. As such, in multi-agent general-sum games, multi-agent IRL algorithms have difficulty balancing cooperative and competitive objectives. To address these issues, we propose Multi-Agent Marginal Q-Learning from Demonstrations (MAMQL), a novel sample-efficient framework for multi-agent IRL.
	%MAMQL uses a per-agent learned marginalization of the action-value function to approximate an agent’s reward function.
	For each agent, MAMQL learns a critic marginalized over the other agents' policies, allowing for a well-motivated use of Boltzmann policies in the multi-agent context. We identify a connection between optimal marginalized critics and single-agent soft-Q IRL, allowing us to apply a direct, simple optimization criterion from the single-agent domain. Across our experiments on three different simulated domains, MAMQL significantly outperforms previous multi-agent methods in average reward, sample efficiency, and reward recovery by often more than \textbf{2-5x}. We make our code available at \href{https://sites.google.com/view/mamql}{https://sites.google.com/view/mamql}.

	%Inverse Reinforcement Learning (IRL) uses expert demonstrations to infer preferences and produce useful policies. In the multi-agent context, the task is complicated by the fact that expert demonstrations must be modeled as equilibria, producing a layer of indirection between ground truth rewards and observed expert behaviors which multi-agent IRL (MAIRL) algorithms must navigate. In this paper, we propose Multi-Agent Marginal Q-Learning from Demonstrations (MAMQL), a novel sample-efficient framework for multi-agent IRL in discrete action spaces. For each agent, MAMQL learns a critic marginalized over the other agents' policies, allowing for a well-motivated use of Boltzmann policies in the multi-agent context. We identify a connection between optimal marginalized critics and single-agent soft-Q IRL, allowing us to apply a direct, simple optimization criterion from the single-agent domain. Across our experiments on three different general-sum domains, MAMQL significantly outperforms previous multi-agent methods in average reward, sample efficiency, and reward recovery by often more than \textbf{2-5x}.
\end{abstract}

\section{Introduction}

Inverse reinforcement learning (IRL) has been widely used to approach many complex problems in control and decision-making, such as autonomous driving \cite{allegra2020inverse,cao2020reinforcement}, resource management \cite{dey2023inverse,wang2023inverse}, and team sports \cite{chen2022reliable,rahimian2021inferring}. These scenarios often involve multiple agents, where each agent must consider both the cooperative and competitive nature of the environment. IRL aims to train agents to learn the optimal policy through the use of expert demonstrations. While effective in single-agent cases, we cannot naively apply single-agent IRL in multi-agent environments because the other agents' strategies influence the policy of the agent we are training. Multi-agent approaches to IRL (MAIRL), such as MA-AIRL \cite{yu2019multi}, MAAC \cite{lowe2017multi}, and MA-DAC \cite{xue2022multiagent}, effectively combat the non-stationarity of multi-agent environments caused by conflicting agent policies. However, they often struggle to robustly recover rewards that explain the expert demonstrations, especially in general-sum games. For example, in autonomous driving, each agent has competitive and cooperative objectives. While the competitive objective of arriving at the destination quickly is clearly defined, cooperating with other agents to avoid collisions in various traffic control systems is more ambiguous. Such an environment has multiple equilibria with agents optimizing for a mix of cooperative/competitive objectives; though straightforward to demonstrate, learning correct objective functions for autonomous driving is unsolved. Thus, we are interested in the following question: how can we learn the agents' objective functions in a multi-agent system that explain the cooperative and competitive behaviors seen in the expert demonstrations?

To solve this problem, we propose \textbf{M}ulti-\textbf{A}gent \textbf{M}arginal \textbf{Q}-\textbf{L}earning from Demonstrations (MAMQL). MAMQL learns a marginalization of the action-value function of each agent by taking the expectation of their value function over the action space of all other agents. MAMQL then uses the marginalized action-value function to approximate a reward function that is representative of the mixed cooperative/competitive behaviors observed in the expert demonstrations.

To test our algorithm, we work with three multi-agent environments of interest. First, we introduce a new general-sum gridworld environment called ``Gems.'' The second is Overcooked \cite{carroll2020utility}, a standard benchmark environment for cooperative multi-agent setups. Third, we utilize the Highway environment \cite{highway-env}, an autonomous driving simulator with scalability for testing the algorithms in scenarios with varying numbers of autonomous vehicles. All three environments provide a rich space of strategies where effective cooperation differs significantly from behavior that does not acknowledge the other agents. Thus, for each environment, the expert demonstrations may contain different equilibria, making reward recovery relatively difficult. Yet, we observe MAMQL is a significant improvement compared to the state-of-the-art methods, illustrating its robustness and sample-efficiency.

\section{Related Work}

Our approach connects ideas from inverse reinforcement learning and imitation learning to exploit the unique structure of multi-agent problems.

\subsection{Inverse Reinforcement Learning}

In traditional reinforcement learning (RL), an agent operating in a Markov decision process (MDP) is trained to maximize the cumulative expected reward by interacting with the environment.
Inverse reinforcement learning (IRL) solves the opposite problem: given expert demonstrations, recover the reward function \cite{ratliff2006maximum, abbeel2004apprenticeship, ng2000algorithms, ramachandran2007bayesian,ho2016generative, waugh2013computational,biyik2022learning}.
After performing IRL, the recovered reward function can be used to train a policy that mimics the expert's behavior.
However, with suboptimal experts, as is typically the case with human data, inferring the true reward becomes increasingly difficult.
One way to address this issue is the MaxEnt IRL framework~\cite{ziebart2008maximum,eysenbach2022maximum} which models suboptimality by merely assuming experts are rational in proportion to the reward they expect to receive.
Another approach, IQ-Learn \cite{garg2021iq}, uses an inverse soft-Q learning framework to directly infer policies from demonstrations.
This reduces the non-stationarity of the environment, streamlining the extraction of reward and policy functions, a property we borrow in MAMQL.
Our contribution is to show that inverse soft-Q learning can be extended to the multi-agent setting by learning marginalized critics for each agent.
A direct extension of IQ-Learn to the multi-agent context serves as a baseline for our algorithm.

\subsection{Imitation Learning}

Broadly, imitation learning (IL) is a technique that learns a policy by mimicking an expert's behavior \cite{brown2019better,ho2016generative,ross2011reduction,sontakke2023roboclip}.
Both IRL and imitation learning (IL) utilize an expert demonstration dataset to learn optimal agent policies, though IRL also infers rewards, potentially at additional cost~\cite{ho2016generative}.

In multi-agent environments, IL encounters difficulties attaining a stable equilibrium during joint optimization \cite{zhang2021imitation}.
Our proposed MAMQL approach, in addition to learning reward functions, learns policies that empirically satisfy a generalized version of the Nash Equilibrium similar to \citet{peters2021inferring}, addressing the limitations of existing IL methods that do not sufficiently capture both cooperative and competitive objectives.
This approach enhances the robustness of learned policies and their similarity to those of expert agents, which is particularly advantageous in scenarios involving strategic interactions among multiple agents.

\subsection{Multi-Agent Learning}
The extension of IRL and IL to the multi-agent case introduces a number of complications.
With more than one agent, the IRL objective becomes to learn a reward function for each agent, while the IL objective requires producing a joint policy as agents' actions depend on their interactions with the other agents, e.g.,
through conventions \cite{shih2020critical,hu2020other}.
The notion of an optimal policy is also much more challenging to define in this context since the expected return of one agent's policy depends on the policies of all other agents.
Extensive non-stationarity hampers theoretical guarantees and causes entire classes of algorithms to exhibit slow convergence to occasionally suboptimal equilibria in practice compared to expert trajectories \cite{lowe2017multi}.
Some techniques, such as MA-AIRL \cite{yu2019multi}, adapt soft methods that model bounded rationality into the multi-agent case, allowing it to connect the significant literature in single-agent MaxEnt IRL \cite{ziebart2008maximum} to the multi-agent case where maximum entropy techniques have seen less exploration.
Our approach adopts a similar framework that interprets an agent's policy as a generalized Boltzmann policy to construct a joint optimization objective.

An additional feature of interest in the multi-agent context is the structure of reward functions.
Zero-sum games, as well as fully cooperative games where all agents share the same reward function, act as simplifying assumptions \cite{zhang2021multiagent}.
Recent work has attempted to tackle multi-agent IRL in such settings \cite{kalogiannis2022efficiently}.
Our algorithm seeks to relax these assumptions by allowing mixed cooperative and competitive objectives, thus dealing with general-sum environments.
In these settings, agents may be balancing complex interactions between incentives, working together toward shared goals in some contexts and competing in others, forcing IRL methods to infer complex equilibria from expert data.

\section{Problem Formulation}

We operate in a Markov game modeled by the tuple $\langle n, \mathcal{S}, \mathcal{A}, T, \mathcal{R}, p_0 \rangle$, where $n$ is the number of agents, $\mathcal{S}$ is the set of states, with each state denoted by \(s \in \mathcal{S}\), $\mathcal{A}$ is a \emph{discrete} set of actions symmetrically available to each agent, $T: \mathcal{S} \times \mathcal{A}^n \rightarrow \Delta(\mathcal{S})$ is the transition function, $\mathcal{R}: \mathcal{S} \times \mathcal{A}^n \rightarrow \mathbb{R}^n$ gives the true rewards for each agent, and $p_0 \in \Delta(\mathcal{S})$ is the distribution of starting states.
For a vector of actions $\bm{a} \in \mathcal{A}^n$, we adopt the notation $(a_i, \bm{a}_{-i})$ to denote a vector of actions where $i^\textrm{th}$ agent's action is $a_i$ and $\bm{a}_{-i}$ represents the actions of all agents except $i$.
We denote a joint policy as $\bm{\pi}$ where $\pi_i: \mathcal{S} \rightarrow \Delta (\mathcal{A})$ is the policy of the $i^\textrm{th}$ agent, and $\bm{\pi}_{-i}$ is the marginal policy of all agents except $i$.

Given the dynamics information $\langle n, \mathcal{S}, \mathcal{A}, T, p_0 \rangle$ without the reward function, and a dataset $\rho_E$ of expert transitions of the form $(s, \bm{a}, s')$, our goal is to reconstruct $\mathcal{R}$. In a general-sum setting, this means we learn a distinct reward function $\mathcal{R}_i$ for each agent \(i\), naturally accommodating cooperative and competitive objectives. If the environment is purely cooperative, $\mathcal{R}_i$ would converge to near-identical functions, whereas in environments with distinct roles or asymmetries, each agent’s reward function may diverge from the others to reflect its unique objectives. Moreover, we preserve agent indices in expert data \((s, \mathbf{a}, s')\) to ensure that each agent \(i\) learns from transitions relevant to its own perspective, which is especially important when agents fulfill different roles (e.g., Overcooked chefs, or vehicles with unique routes in a traffic network). Although one could shuffle indices in perfectly symmetric tasks, real-world domains typically exhibit role-specific state and action distributions, making fixed indexing the more natural approach.

\section{Approach}

%%VM.5.20: TODO: look at best response equilibrium analysis from MA-AIRL paper

%We consider a multi-agent setup where each agent takes simultaneous actions and receives rewards based on their individual reward functions, formalized as a tuple $\langle n, \mathcal{S}, \mathcal{A}, T, \mathcal{R}, p_0 \rangle$. In each timestep $t$, all agents observe the game state $s \in S$, take actions according to their respective policies, and receive rewards according to their individual reward functions $R_i$, assuming they know each other’s objectives and policies.

%\subsection{Background}

We develop our approach by studying the behavior of Boltzmann policies in a multi-agent equilibrium.
Yu et al.~\citet{yu2019multi} argue that joint policies which satisfy a desirable optimality condition, where each $\pi_i$ is the optimal response modulated by a rationality parameter $\lambda > 0$ to fixed $\bm{\pi}_{-i}$, must satisfy the recurrence
\begin{align}
	\pi_i(a_i \!\mid\! s) \!=\! \frac{\exp \lambda \mathbb{E}_{\tilde{\bm{a}}_{-i} \sim \bm{\pi}_{-i}(\cdot \mid s)} [Q_i^{\bm{\pi}}(s, a_i, \tilde{\bm{a}}_{-i})]}{\sum_{\tilde{a}_i \in \mathcal{A}}\! \exp \lambda \mathbb{E}_{\tilde{\bm{a}}_{-i} \sim \bm{\pi}_{-i}(\cdot \mid s)} [Q_i^{\bm{\pi}}(s, \tilde{a}_i, \tilde{\bm{a}}_{-i})]}
	\label{eq:recurrence}
\end{align}
for all state-action pairs, where $Q_i^{\bm{\pi}}$ is the true expected discounted return of $\pi_i$ given fixed $\bm{\pi}_{-i}$. Given a learned critic $\hat{Q}_i^{\bm{\pi}}: \mathcal{S} \times \mathcal{A}^n \rightarrow \mathbb{R}$ which estimates expected return for agent $i$, a policy $\pi_i$ satisfying \cref{eq:recurrence} for $\hat{Q}_i^{\bm{\pi}}$ in place of $Q_i^{\bm{\pi}}$ can be interpreted as a ``generalized Boltzmann policy'' with respect to the critic.

As a single-agent IRL method, IQ-Learn \cite{garg2021iq} assumes expert policies to be generalized Boltzmann policies with respect to a learned critic. With this assumption, it produces a non-adversarial optimization objective, unlike previous methods involving adversarial optimization like GAIL \cite{ho2016generative}. We follow a similar path with generalized Boltzmann policies to construct our algorithms in the multi-agent context.

\subsection{Multi-Agent Marginal Q-Learning (MAMQL)}

Let us first consider a learning setup where, for each agent \(i\), we learn a critic \(Q_{\psi_i}: \mathcal{S} \times \mathcal{A}^n \rightarrow \mathbb{R}\). If we wish to compute \(\bm{\pi}\) as a generalized Boltzmann policy given the critics, the form of \cref{eq:recurrence} does not provide an explicit construction, as \(\pi_i(a_i \mid s)\) depends on \(\bm{\pi}_{-i}(\bm{a}_{-i} \mid s)\). Although it is possible to learn policies alongside critics and enforce \cref{eq:recurrence} through joint optimization, we simplify the setup by learning \emph{marginalized} critics \(\bar{Q}_{\psi_i} : \mathcal{S} \times \mathcal{A} \rightarrow \mathbb{R}\). That is, for a fixed \(\bm{\pi}_{-i}\), we require that the marginalized critic is defined as the expectation over \(\tilde{\bm{a}}_{-i}\) sampled according to \(\bm{\pi}_{-i}(\cdot \mid s)\), i.e., 
\begin{align}
	\bar{Q}_{\psi_i}(s, a_i) = \mathbb{E}_{\tilde{\bm{a}}_{-i} \sim \bm{\pi}_{-i}(\cdot \mid s)}[Q^{\pi_i}_i(s, a_i, \tilde{\bm{a}}_{-i})]\:.
	\label{eq:critic_condition}
\end{align}
A set of policies satisfying \cref{eq:recurrence} must be Boltzmann policies of a set of marginalized critics $\bar{Q}_{\psi_i}$ satisfying \cref{eq:critic_condition} by direct substitution. Thus, we will henceforth treat all policies $\bm{\pi}$ as Boltzmann with respect to a set of marginalized critics so that
\begin{align}
	\pi_i(a_i \mid s) = \frac{\exp \lambda \bar{Q}_{\psi_i}(s, a_i)}{\sum_{a'} \exp \lambda \bar{Q}_{\psi_i}(s, a')}\:.
	\label{eq:boltzmann}
\end{align}

Next, we generalize the soft Bellman condition given by Haarnoja et al.~\citet{haarnoja2018soft} to the multi-agent setting. We take the soft value function for agent $i$ given $\bm{\pi}$ as
\begin{align}
	V^{\bm{\pi}}_i(s) = \mathbb{E}_{\tilde{\bm{a}} \sim \bm{\pi}(\cdot \mid s)} [Q_i^{\bm{\pi}}(s, \tilde{\bm{a}}) - \log \pi_i(\tilde{\bm{a}}_i \mid s)]
	\label{eq:soft_value_function}
\end{align}
where we note that the second term inside the expectation penalizes the entropy of $\pi_i$ only, rather than $\bm{\pi}$. We substitute \cref{eq:boltzmann} in \cref{eq:soft_value_function} and simplify to get:
\begin{align}
	V^{\bm{\pi}}_i(s) = V^{\bm{\pi}}_{\psi_i}(s) : & = (1-\lambda) \mathbb{E}_{a_i \sim \pi_i(\cdot \mid s)}[\bar{Q}_{\psi_i}(s, a_i)  \nonumber \\
	                                               & \qquad + \log \sum_{a'_i \in \mathcal{A}} \exp \lambda \bar{Q}_{\psi_i}(s, a'_i).
	\label{eq:value_est}
\end{align}

We note a connection between \cref{eq:recurrence} and maximum entropy reinforcement learning\footnote{See Section 3.1 of \citet{yu2019multi} for relevant discussion in a fixed-horizon setting.}: for fixed $\bm{\pi}_{-i}$, the choice of $\bar{Q}_{\psi_i}$ which makes the Boltzmann policy $\pi_i$ satisfy \cref{eq:recurrence} must also satisfy a marginalized soft Bellman condition:
\begin{align}
	 & \mathbb{E}_{\tilde{\bm{a}}_{-i} \sim \bm{\pi}_{-i}(\cdot \mid s)}[\mathcal{R}_i(s, a_i, \tilde{\bm{a}}_{-i})] \!=\! \bar{R}^{\bm{\pi}}_{\psi_i}(s, a_i) \!=\! \bar{Q}_{\psi_i}(s, a_i) \nonumber \\
	 & \qquad \quad - \gamma \mathbb{E}_{\tilde{\bm{a}}_{-i} \sim \bm{\pi}_{-1}(\cdot \mid s), s' \sim T(\cdot \mid s, a_i, \tilde{\bm{a}}_{-i})}[V^{\bm{\pi}}_{\psi_i}(s')].  \label{eq:reward_est}
	%%VM.5.07: ":=" means "is defined as"
	%%VM.5.07: these compound statement with "=" and ":=" are a bit confusing, could we break them up?
	%%NH: as in work out the subsitution of πi into Vi to get Vψ? That seems a bit verbose arithmetic or am I missing something?
\end{align}

For a set of marginalized critics satisfying \cref{eq:critic_condition}, all choices of $s$ and $a_i$ in \cref{eq:reward_est} introduce $\abs{\mathcal{S}} \abs{\mathcal{A}}$ constraints on $\mathcal{R}_i$. Namely, a learned reward estimate $R_{\mu_i}$ for $\mathcal{R}_i$ must, for each $(s, a_i)$ pair and fixed marginal critics, satisfy the constraint
\begin{align}
	\bar{R}^{\bm{\pi}}_{\psi_i}(s, a_i) = \mathbb{E}_{\tilde{\bm{a}}_{-i} \sim \bm{\pi}_{-i}(\cdot \mid s)} R_{\mu_i}(s, a_i, \tilde{\bm{a}}_{-i})\:.
	\label{eq:reward_constraint}
\end{align}

Let us trace back through these results to emphasize their role. We found that joint policies satisfying desirable optimality conditions \cref{eq:recurrence} can be expressed with the Boltzmann policies of a set of marginalized critics. The non-stationarity of expected return for an agent's policy (since it depends on other agents' policies) is taken into account implicitly by the definition of the marginalized critics (\cref{eq:critic_condition}), allowing for a simple expression of the joint policy $\bm{\pi}$ in terms of the Boltzmann policies of the learned critics (\cref{eq:boltzmann}). Each marginal critic is also an optimal soft-Q function in the single-agent MDP induced by fixed $\bm{\pi}_{-i}$, since it satisfies \cref{eq:reward_est}, which is equivalent to the soft Bellman condition for that MDP. The marginalized critics provide a set of conditions which the experts' reward functions must meet (\cref{eq:reward_constraint}).

\subsection{Training}

The marginal $Q$-function $\bar{Q}_{\psi_i}$ is an optimal soft-Q function in the single-agent MDP induced by fixed $\bm{\pi}_{-i}$. However, identifying an optimal soft $Q$-function whose Boltzmann policy best satisfies an expert dataset in the single-agent setting is exactly the question considered by Garg et al.~\citet{garg2021iq}. In the discrete setting, the authors show that the optimal soft $Q$-function is obtained by a concave objective in terms of the learned $Q$-function and its soft value estimate. Translating their objective to our setting gives the critic loss function for agent $i$:
\begin{equation}
	\min_{\psi_i} (1 - \gamma) \mathbb{E}_{s_0 \sim p_0}[V^{\bm{\pi}}_{\psi_i}(s_0)] - \mathbb{E}_{(s, \bm{a}) \sim \rho_E}[\phi(\bar{R}^{\bm{\pi}}_{\psi_i}(s, \bm{a}_i))]
\end{equation}
where $\rho_E$ is the state-action distribution of the expert dataset, and $\phi$ is a concave reward regularizer. Several examples of regularizers in terms of common divergence functions are given in Appendix B of \citet{garg2021iq}. For example, $\phi(x) = x$ corresponds to total variation divergence while $\phi(x) = x - \frac{x^2}{4}$ corresponds to Pearson $\chi^2$ divergence. It is the latter regularizer we use in our experiments.

This objective allows for offline learning, but, following the discussion by Garg et al.~\citet{garg2021iq}, Section 5.1, we achieve better performance in an online version by sampling $\mathbb{E}_{(s, \bm{a}, s') \sim \rho_\pi}[V^{\bm{\pi}}(s) - \gamma V^{\bm{\pi}}(s')]$ to substitute $(1 - \gamma) \mathbb{E}_{s_0 \sim p_0}[V^{\bm{\pi}}_{\psi_i}(s_0)]$ in our loss function, where $\rho_\pi$ is the distribution of transitions induced by joint policy $\bm{\pi}$.

The learned marginalized critics do not directly provide estimates of rewards for individual state-action pairs. However, \cref{eq:reward_constraint} shows that we recover a set of constraints for $R_{\mu_i}$. To reconstruct $R_{\mu_i}$, we sample state-action pairs from $\rho_\pi$ and update according to an MSE objective which enforces those constraints:
\begin{align}
	 & \min_{\mu_i} \mathbb{E}_{(s, \bm{a}) \sim \rho_\pi} \Bigl[ \bigl( \bar{R}^{\bm{\pi}}_{\psi_i}(s, \bm{a}_i) \nonumber                                                            \\
	 & \qquad - \mathbb{E}_{\tilde{\bm{a}}_{-i} \sim \bm{\pi}_{-i}(\cdot \mid s)} R_{\mu_i}(s, \bm{a}_i, \tilde{\bm{a}}_{-i}) \bigr)^2 \Bigr] + \beta(R_{\mu_i}) \label{eq:reward_opt}
\end{align}
where $\beta$ is a reward regularization component. The pseudocode of the online method is provided in \cref{alg:mamql}.

One practical consideration is that it is often difficult, given an expert transition $(s, \bm{a}, s')$, to exactly compute the term $\mathbb{E}_{\tilde{\bm{a}}_{-i} \sim \bm{\pi}_{-1}(\cdot \mid s), s'' \sim T(\cdot \mid s, a_i, \tilde{\bm{a}}_{-i})}[V^{\bm{\pi}}_{\psi_i}(s'')]$ from \cref{eq:reward_est}, as it in general requires computing $\abs{\mathcal{A}}^{n-1}$ transitions. Even unbiased sampling is unwieldy, requiring importance sampling that leverages state-action occupancy measures from the expert dataset. We found an acceptable alternative is the biased one-sample estimate of the term as $V^{\bm{\pi}}_{\psi_i}(s')$, since intuitively the expert joint actions $\bm{a}_{-i}$ which resulted in $s'$ should be likely to be sampled by $\bm{\pi}_{-i}$ once it is performing well enough. This is the approach we employ in our experiments, which we will present next.

\begin{algorithm}
	\caption{\setup}
	\label{alg:mamql}
	\begin{algorithmic}[1]
		%%VM.5.07: why does n need to be part of the environment? it would be good to be standard with the notation (gamma can be part of the MDP as well)
		%%NH: N isn't necessary, I belive we can remove it and treat s,a,s' as vectors, would that work?
		\State {\bfseries Input:} \parbox[t]{\linewidth}{%
		Environment $\langle n, \mathcal{S}, \mathcal{A}, T, p_0 \rangle$ \\
		Expert transitions $\rho_E = \{(s_{i}, \bm{a}_{i}, s'_{i})\}_{i=1}^{N}$ \\
		Buffer $\rho_\pi$ of agents' rollout transitions \\
		Regularization function $\phi$ \\
		Discount factor $\gamma$, learning rate $\alpha$%
		}
		\State Initialize $\psi_i$, for marginal critics $\bar{Q}_{\psi_i}$ and each agent $i$
		\State Initialize $\mu_i$, for reward function $R_{\mu_i}$ and each agent $i$
		\Repeat
		\State Add rollout trajectory to $\rho_\pi$ using joint policy $\bm{\pi}_{\bm{\psi}}$
		\For{$i=1$ {\bfseries to} $N$}
        \State\parbox[t]{\linewidth}{%
            Sample expert and rollout transitions \\[-1pt]
            \mbox{\hspace*{6ex}$\bm{\chi}_E \sim \rho_E$ and $\bm{\chi}_\pi \sim \rho_\pi$}%
            \medskip
        }
        \State \parbox[t]{\linewidth}{%
            \hbox{%
                $\psi_i \gets \psi_i - \alpha\nabla_{\psi_i}\bigl(\mathbb{E}_{\bm{\chi}_\pi}
                [V^{\bm{\pi}}_{\psi_i}(s)- \gamma V^{\bm{\pi}}_{\psi_i}(s')]\hfill$%
            }
            \hbox to \linewidth{%
                \hspace*{10ex}
                $- \mathbb{E}_{\bm{\chi}_E}[\phi(\bar{R}^{\bm{\pi}}_{\psi_i}(s, \bm{a}_i))]\bigr)$
                \hfill (\ref{eq:value_est}, \ref{eq:reward_est})%
            }%
        }
		\State \parbox{\linewidth}{Update $\mu_i$ using the sample mean over $\bm{\chi}_\pi$ \hfill (\ref{eq:reward_opt})}
		\EndFor
		\Until convergence
		\State \Output Learned policies $\bm{\pi}_{\bm \psi}$ and reward functions $\bm{R}_{\bm{\mu}}$
		% (and if performing IRL, learned reward functions $\bm{R}_{\bm{\mu}}$)
	\end{algorithmic}
\end{algorithm}

\section{Experiments}

\begin{figure*}[htb!]
	\centering
	\includegraphics[width=0.9\linewidth]{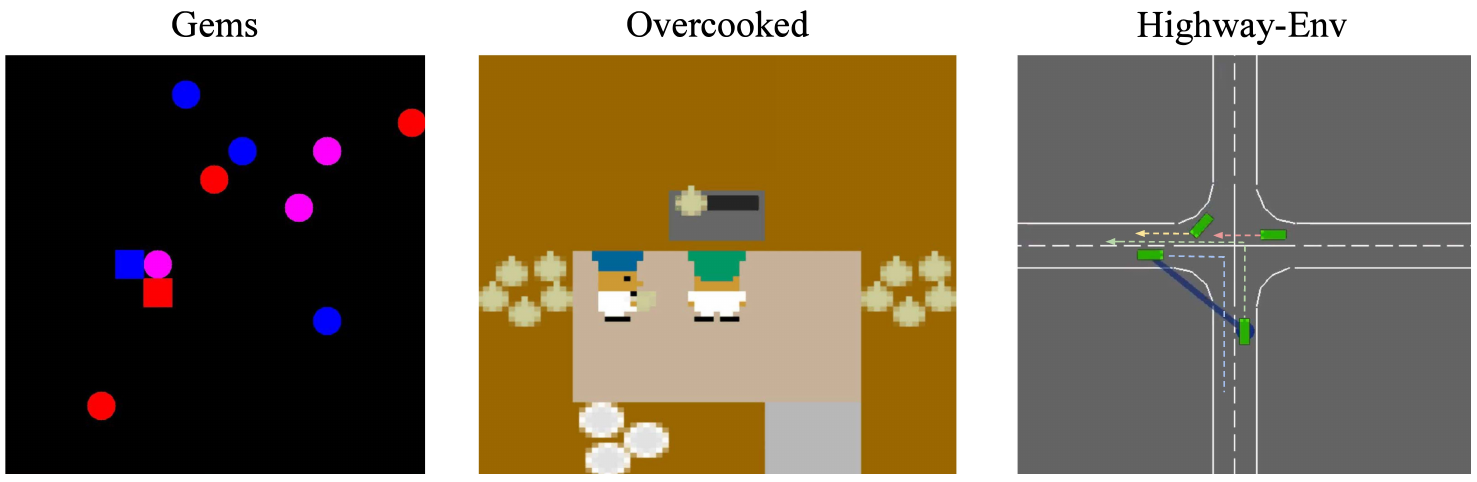}
	\caption{On the left, the proposed multi-agent grid world environment where above square agents are working together to get a purple circular gem. At the center, the Overcooked cramped environment with one agent adding onions to the soup and the other agent ready with another. On the right, a four agent intersection from the bottom agent's viewpoint. The weighted line marks priority level of collision avoidance with another agent.}
    \vspace{-10px}
	\label{fig:envs}
\end{figure*}

To test the performance of MAMQL, we conducted experiments on three different simulated multi-agent environments. We will first introduce those environments in the next subsection. We provide visuals of the environments in \cref{fig:envs}.

\subsection{Environments}

As our simplest testbed, we introduce a new environment called ``Gems,'' a grid world where two agents (Red and Blue) collect gems. The gems are also red and blue and can only be collected by the agent of the same color and give $1$ a reward. Additionally, there are purple gems, which are $6 \times$ as rewarding and can be collected by both agents, but only when they are standing on (possibly distinct) purple gems simultaneously. The game has a horizon of $45$ moves. The action space consists of moving one step in the four cardinal directions and a stopping action.

We also performed experiments in the Overcooked environment \cite{carroll2020utility, knott2021evaluating, ribeiro2022assisting, zhao2022maximum}. This environment has two agents who can move and interact with various ingredients and cooking tools. Each agent receives a reward of $10$ when anyone delivers finished onion soup to a customer. Each agent's action space involves moving in each cardinal direction, interacting with objects, or stopping. Overcooked has a maximum time horizon of $50$ steps.

Finally, we performed experiments in the multi-agent intersection scenario from Highway-Env \cite{chen2022deep, chen2020delayaware, highway-env, n2020smart}. We modified the original environment with only two agents to test our algorithm's ability to learn mixed cooperative and competitive behaviors in traffic with many agents. Specifically, our modified environment consists of four agents. Each agent can decrease or increase their velocity to successfully negotiate right-of-way at a four-way intersection to arrive at a target destination without collision. Each agent receives a reward of $2$ for arriving at their individual destination and a reward of $-4$ for colliding with another agent(s). Each agent's action space involves decreasing, maintaining, or increasing velocity. An episode concludes when an agent crashes, or all agents reach their destination. While the agents are expected to demonstrate cooperation to avoid collisions, the discount factor introduces competition between them to get the right-of-way.

\subsection{Baselines}

We compare MAMQL against three alternative baselines. The first one is behavioral cloning where we learn a policy for each agent separately. Similarly, our second baseline is IQ-Learn \cite{garg2021iq} where we again learn the policies and the reward functions separately for different agents.

We also implemented a direct multi-agent extension of IQ-Learn, as the ``IQ-Learn MA'' baseline, by expanding the type signature of agent $i$'s critic to include all agents actions (i.e., from type signature $\mathcal{S} \times \mathcal{A} \rightarrow \mathbb{R}$ of a single-agent critic to $\mathcal{S} \times \mathcal{A}^n \rightarrow \mathbb{R}$ which uses all agent actions from sampled transitions, but otherwise follows the IQ-Learn implementation exactly). Lastly, we compare MAMQL against the MA-AIRL \cite{yu2019multi} algorithm.

\subsection{Datasets}
For the Gems environment, we built an expert dataset from rollouts of expert RL policies trained using MADDPG \cite{lowe2017multi}, with a CNN architecture consisting of $3$ convolutional layers with an MLP head with $3$ hidden layers, with batch normalization between convolutional layers and a LeakyReLU activation function. To help train the experts, we also introduced reward shaping to encourage agents to approach purple gems. For Overcooked, we used a preexisting RL agent with an LSTM+MLP architecture \cite{carroll2020utility} to produce our expert dataset. For the multi-agent Highway-Env scenario we trained an expert RL policy using DQN with social attention \cite{leurent2019social}.

For each environment, we constructed several datasets of simulated expert transitions ranging from $500$ to $70,\!000$ environment steps.

\subsection{Implementation Details}

For the Gems environment, we ran all methods by training CNN architectures that have $3$ convolutional layers and an MLP head with 3 layers; ELU is used as the activation function and modest dropout weakly regularizes the networks. For the Overcooked and Highway-Env environments, we tested all our methods using an MLP architecture with four hidden layers and ELU as the activation function.

We comprehensively tuned available hyperparameters (regularization parameters, buffer size, number of expert examples, etc.) and report the average episodic return for each algorithm, each averaged across $1000$ trajectories. Specifically, for a trajectory $\tau$ of length $T$ in an environment with $N$ agents, we define the episodic return $G(\tau)$ as
\begin{align}
    G(\tau) \;=\; \sum_{t=0}^{T-1}\,\sum_{i=1}^N r_t^i
    \label{eq:episodic_return}
\end{align}
where $r_t^i$ is the reward received by agent $i$ at time step $t$. The reported value is the mean of $G(\tau)$ over the $1000$ test rollouts. All experiments were run on a computing cluster consisting of eight RTX A6000s; however, we note \textbf{MAMQL} can be trained on a single RTX 3060 with 6\,GB GDDR6.

\section{Results}

%%VM.5.07: general comment: use descriptive names for labels, the numbers may change with layout
Each entry of \cref{tab:avg_reward} shows the average return of the agents for the Overcooked and Highway-Env environments across 1000 episodes. The expert dataset size for all algorithms for Gems, Overcooked, and Highway-Env was 20k, 2k, 70k episodes respectively.
We additionally plot the ratio of returns,
[Agent 1 average return]/[Agent 2 average return],
for the Gems environment since it is possible for the agents to acquire considerably different returns there.

\begin{table}[htb!]
	\caption{Average Reward}
	\centering
	\begin{tabular}{r | c c c}
		\toprule
		Algorithm      & Gems                 & Overcooked     & Highway-Env   \\
		\midrule
		Expert         & 14.42/14.42          & 24.3           & 1.72          \\
		BC             & 4.55/4.35            & 0.01           & -0.14         \\
		IQ-Learn Indp  & 8.65/8.65            & 0              & -1.14         \\
		IQ-Learn MA    & 11.45/11.45          & 8.48           & 0.31          \\
		MA-AIRL        & 2.04/2.04            & 0.01           & -1.62         \\
		\textbf{MAMQL} & \textbf{13.05/13.05} & \textbf{13.23} & \textbf{1.71} \\
		\bottomrule
	\end{tabular}
	\label{tab:avg_reward}
\end{table}

Comparing \setup\ with the baselines, we found that \setup\ outperformed all baselines approaching or matching the average reward of the expert in the cases of the Gems and Highway-Env environments. Additionally, \setup\ proved to converge roughly $4\times$ faster than IQ-Learn MA.

On the Gems environment, \setup\ learns a policy on par with the expert and is qualitatively the only algorithm we observed to consistently balance cooperative and competitive rewards. On average, the policy learned by IQ-Learn MA only targets purple gems while \setup\ agents balance shared rewards with individual rewards, deviating to pick up red or blue gems on their way to the next purple gem. In the Overcooked environment, \setup\ is able to consistently complete a three-stage task, taking approximately $50$ steps in the correct order to complete. Imitation learning and inverse reinforcement learning using single-agent methods (behavioral cloning and IQ-Learn) never successfully completed the task. On the multi-agent Highway-Env environment, we tested \setup's ability to scale to several agents. With a continuous state space and discrete action space we trained $4$ agents to negotiate right-of-way for a four-way intersection. \setup\ was able to achieve approximately expert-level average rewards with other MAIRL algorithms often unable to avoid collision in this challenging environment.

\begin{table}[htb!]
	\caption{Average Number of Episodes in Thousands for Convergence}
	\centering
	\begin{tabular}{r | c c c}
		\toprule
		Algorithm      & Gems        & Overcooked & Highway-Env \\
		\midrule
		IQ-Learn MA    & 45          & 31         & 43          \\
		MA-AIRL        & 100         & 100        & 100         \\
		\textbf{MAMQL} & \textbf{13} & \textbf{4} & \textbf{10} \\
		\bottomrule
	\end{tabular}
	\label{tab:convergence_time}
\end{table}

\long\def\makeplots#1{
    \addplot [theme1] table [x=step, y=MAMQL] {#1};%
    \addlegendentry{MAMQL}%
    \addplot [theme2] table [x=step, y=MA-AIRL] {#1};%
    \addlegendentry{MA-AIRL}%
    \addplot [text5] table [x=step, y=IQL-MA] {#1};%
    \addlegendentry{IQL-MA}%

    \addplot [name path=upper,draw=none,no markers] table [x=step, y expr=\thisrow{MAMQL}+\thisrow{MAMQL_ERR}] {#1};
    \addplot [name path=lower,draw=none,no markers] table [x=step, y expr=\thisrow{MAMQL}-\thisrow{MAMQL_ERR}] {#1};
    \addplot [fill=theme1,opacity=0.2] fill between[of=upper and lower];

    \addplot [name path=upper,draw=none,no markers] table [x=step, y expr=\thisrow{MA-AIRL}+\thisrow{MA-AIRL_ERR}] {#1};
    \addplot [name path=lower,draw=none,no markers] table [x=step, y expr=\thisrow{MA-AIRL}-\thisrow{MA-AIRL_ERR}] {#1};
    \addplot [fill=theme2,opacity=0.2] fill between[of=upper and lower];

    \addplot [name path=upper,draw=none,no markers] table [x=step, y expr=\thisrow{IQL-MA}+\thisrow{IQL-MA_ERR}] {#1};
    \addplot [name path=lower,draw=none,no markers] table [x=step, y expr=\thisrow{IQL-MA}-\thisrow{IQL-MA_ERR}] {#1};
    \addplot [fill=text5,opacity=0.2] fill between[of=upper and lower];
}

\pgfplotstableread{
step          0   1000   10000   20000
IQL-MA        0   2.38   7.3     11.45
MAMQL         0   8.55   12.75   13.05
MA-AIRL       0   0.63   0.85    2.04
IQL-MA_ERR    0   0.21   0.64    1.01
MAMQL_ERR     0   0.75   1.12    1.15
MA-AIRL_ERR   0   0.06   0.07    0.18
}\gemsdata

\pgfplotstableread{
step          0   500     1000    2000
IQL-MA        0   6.48    7.91    8.48
MAMQL         0   12.81   12.97   13.23
MA-AIRL       0   0       0       0.01
IQL-MA_ERR    0   0.23    0.29    0.31
MAMQL_ERR     0   0.46    0.47    0.48
MA-AIRL_ERR   0   0.01    0.01    0.01
}\overcookeddata

\pgfplotstableread{
step          0   5000    20000   35000   50000   70000
IQL-MA        0   -2.02   -1.9    -1.65   0.06    0.31
MAMQL         0   1.49    1.6     1.68    1.67    1.71
MA-AIRL       0   -2.85   -2.37   -2.23   -1.91   -1.62
IQL-MA_ERR    0   0.40    0.38    0.33    0.01    0.06
MAMQL_ERR     0   0.30    0.32    0.34    0.33    0.34
MA-AIRL_ERR   0   0.57    0.47    0.45    0.38    0.32
}\highwaydata

\pgfplotstabletranspose[colnames from=step,input colnames to=step]\gemsdata{\gemsdata}
\pgfplotstabletranspose[colnames from=step,input colnames to=step]\overcookeddata{\overcookeddata}
\pgfplotstabletranspose[colnames from=step,input colnames to=step]\highwaydata{\highwaydata}

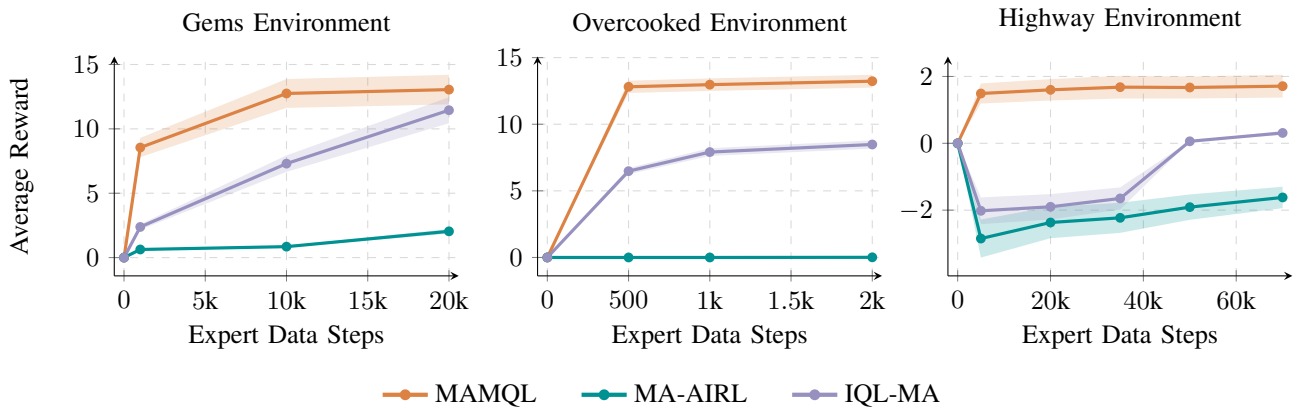
\begin{figure*}[htb!]%
	\centering%
    \pgfplotsset{%
        xticklabel={%
            \pgfmathparse{\tick}%
            \pgfmathtruncatemacro\trunc{\pgfmathresult}%
            \pgfkeys{/pgf/fpu=true}%
            \ifnum\trunc<1000%
                \pgfmathprintnumber{\trunc}%
            \else%
                \pgfmathparse{\pgfmathresult/1000}%
                \ifnum\trunc<1000000\relax%
                    \pgfmathprintnumber{\pgfmathresult}k%
                \else%
                    \pgfmathparse{\pgfmathresult/1000}%
                    \pgfmathprintnumber{\pgfmathresult}M%
                \fi%
            \fi%
            \pgfkeys{/pgf/fpu=false}%
        },%
        grid=major,%
        grid style={gray!30,dashed},%
        width=1.05\linewidth,%
        height=4.5cm,%
        legend cell align=left,%
        axis lines=left,%
        scaled x ticks=false,%
        legend style={draw=none,at={(0.79,0.48)},anchor=center,/tikz/every even column/.append style={column sep=0.5cm},legend image post style={ultra thick}},%
        legend columns=-1,%
        enlarge x limits=0.03,%
        enlarge y limits=0.1,%
    }%
    \hfill%
    \begin{minipage}[t]{0.33\linewidth}%
        \begin{tikzpicture}%
            \begin{axis}[%
                title={Gems Environment},%
                legend to name=leg:plot,%
                name=plot,%
                xlabel={Expert Data Steps},%
                ylabel={Average Reward},%
                every axis plot/.style={very thick,mark=*,mark size=1.25pt},%
            ]%
                \makeplots\gemsdata%
            \end{axis}%
        \end{tikzpicture}%
    \end{minipage}%
    \hspace*{3ex}
    \begin{minipage}[t]{0.33\linewidth}%
        \begin{tikzpicture}%
            \begin{axis}[%
                title={Overcooked Environment},%
                name=plot,%
                legend to name=leg:plot2,%
                xlabel={Expert Data Steps},%
                every axis plot/.style={very thick,mark=*,mark size=1.25pt},%
            ]%
                \makeplots\overcookeddata%
            \end{axis}%
        \end{tikzpicture}%
    \end{minipage}%
    \hspace*{-4ex}
    \begin{minipage}[t]{0.33\linewidth}%
        \begin{tikzpicture}%
        \begin{axis}[%
            title={Highway Environment},%
            name=plot,%
            xlabel={Expert Data Steps},%
            legend to name=leg:plot3,%
            every axis plot/.style={very thick,mark=*,mark size=1.25pt},%
        ]%
            \makeplots\highwaydata% 
        \end{axis}%
        \end{tikzpicture}%
    \end{minipage}%
    \hfill%
    \vbox{\medskip\ref*{leg:plot}}%
    \caption{Reward of recovered policy for MAMQL and baselines, across varying dataset sizes.}%
	\label{fig:baselines}%
\end{figure*}

We also measure convergence time across our environments in \cref{tab:convergence_time} as the average number of episodes (in thousands) to converge to a good policy. The expert dataset size for all algorithms for Gems, Overcooked, and Highway-Env was 20k, 2k, 70k episodes respectively. \setup\ converges $2$-$4\times$ faster to a near-optimal policy across all environments compared to our baselines. Additionally, while \setup\ at most requires $14.5$k episodes to converge to an optimal policy, other MAIRL algorithms generally take up to $100$k episodes to train a good policy on environments of similar complexity to Gems, Overcooked, and Highway-Env based on previous work \cite{jeon2020scalable}.

To distinguish our method as sample efficient instead of quickly overfitting to a dataset that encompasses the majority of the state/action space of the environment, we downscale our dataset size for each environment and observe the average reward over $1000$ episodes of the learned policy (\cref{fig:baselines}). For each environment, \setup\ outperforms our baselines in average reward with a significantly slower performance decay rate as the dataset size decreases.

\pgfplotstableread{
step          0       1000    2000    3000    4000    5000    6000    7000    8000    9000    10000   11000   12000   13000   14000   15000   16000   17000   18000   19000   20000
MAMQL         44.25   3.22    3.13    2.03    1.91    1.88    1.83    1.79    1.79    1.71    1.62    1.66    1.70    1.69    1.65    1.55    1.59    1.61    1.58    1.57    1.50
IQL-MA        45.41   25.56   15.79   12.08   9.70    8.11    6.85    5.91    5.25    4.64    4.18    3.85    3.68    3.57    3.47    3.44    3.40    3.41    3.43    3.39    3.39
MA-AIRL       41.09   41.51   41.22   41.14   41.09   41.09   40.83   40.58   40.50   40.40   40.35   40.28   39.93   39.58   39.04   38.96   38.90   38.81   37.68   37.63   35.59
MAMQL_ERR     0.46    0.03    0.03    0.02    0.02    0.02    0.02    0.02    0.02    0.02    0.02    0.02    0.02    0.02    0.02    0.02    0.02    0.02    0.02    0.02    0.02
IQL-MA_ERR    0.48    0.27    0.17    0.13    0.10    0.09    0.07    0.06    0.06    0.05    0.04    0.04    0.04    0.04    0.04    0.04    0.04    0.04    0.04    0.04    0.04
MA-AIRL_ERR   0.43    0.44    0.43    0.43    0.43    0.43    0.43    0.43    0.43    0.42    0.42    0.42    0.42    0.42    0.41    0.41    0.41    0.41    0.40    0.40    0.37
}\gemsdata

\pgfplotstableread{
step          0          1000        2000       3000       4000       5000       6000       7000      8000      9000      10000  
MAMQL         31962.62   108.27      71.45      46.76      45.91      44.19      39.81      7.10      1.24      1.98      1.60
IQL-MA        31470.23   3799.00     2633.93    1858.98    1593.28    1122.61    792.78     668.63    573.47    477.66    463.29
MA-AIRL       31069.12   3224.95     1830.74    1181.13    1128.24    1452.33    1426.80    1500.98   1443.35   1570.63   1732.05
MAMQL_ERR     2.71       0.01        0.01       0.00       0.00       0.00       0.00       0.00      0.00      0.00      0.00
IQL-MA_ERR    2.67       0.32        0.22       0.16       0.14       0.10       0.07       0.06      0.05      0.04      0.04
MA-AIRL_ERR   2.64       0.27        0.16       0.10       0.10       0.12       0.12       0.13      0.12      0.13      0.15
}\overcookeddata

\pgfplotstableread{
step          0          1000        2000       3000       4000       5000       6000       7000      8000      9000      10000     11000     12000     13000     14000     15000     16000     17000     18000     19000     20000
MAMQL         96.82      1.32        1.18       1.20       1.18       1.17       1.11       1.14      1.10      1.07      1.14      1.08      1.11      1.26      1.13      1.27      1.26      1.16      1.12      1.08      1.02
IQL-MA        98.74      104227.95   68367.82   37730.12   21759.85   14302.71   11306.50   7466.48   6222.10   3257.58   3146.55   2224.22   1254.02   930.61    472.73    631.96    399.87    361.21    324.93    332.92    313.79
MA-AIRL       97.01      2060.73     1937.75    1919.28    1930.22    1919.28    1900.55    1889.78   1886.54   1887.79   1876.49   1862.46   1856.19   1850.32   1860.56   1843.70   1799.34   1793.89   1765.08   1642.51   1502.74
MAMQL_ERR     0.01       0.00        0.00       0.00       0.00       0.00       0.00       0.00      0.00      0.00      0.00      0.00      0.00      0.00      0.00      0.00      0.00      0.00      0.00      0.00      0.00
IQL-MA_ERR    0.01       7.47        4.90       2.70       1.56       1.02       0.81       0.53      0.45      0.23      0.23      0.16      0.09      0.07      0.03      0.05      0.03      0.03      0.02      0.02      0.02
MA-AIRL_ERR   0.01       0.15        0.14       0.14       0.14       0.14       0.14       0.14      0.14      0.14      0.13      0.13      0.13      0.13      0.13      0.13      0.13      0.13      0.13      0.12      0.11
}\highwaydata

\pgfplotstabletranspose[colnames from=step,input colnames to=step]\gemsdata{\gemsdata}
\pgfplotstabletranspose[colnames from=step,input colnames to=step]\overcookeddata{\overcookeddata}
\pgfplotstabletranspose[colnames from=step,input colnames to=step]\highwaydata{\highwaydata}

\begin{figure*}[htb!]%
	\centering%
    \pgfplotsset{%
        xticklabel={%
            \pgfmathparse{\tick}%
            \pgfmathtruncatemacro\trunc{\pgfmathresult}%
            \pgfkeys{/pgf/fpu=true}%
            \ifnum\trunc<1000%
                \pgfmathprintnumber{\trunc}%
            \else%
                \pgfmathparse{\pgfmathresult/1000}%
                \ifnum\trunc<1000000\relax%
                    \pgfmathprintnumber{\pgfmathresult}k%
                \else%
                    \pgfmathparse{\pgfmathresult/1000}%
                    \pgfmathprintnumber{\pgfmathresult}M%
                \fi%
            \fi%
            \pgfkeys{/pgf/fpu=false}%
        },%
        grid=major,%
        grid style={gray!30,dashed},%
        width=1.05\linewidth,%
        height=4.5cm,%
        legend cell align=left,%
        axis lines=left,%
        scaled x ticks=false,%
        legend style={draw=none,at={(0.79,0.48)},anchor=center,/tikz/every even column/.append style={column sep=0.5cm},legend image post style={ultra thick}},%
        legend columns=-1,%
        enlarge x limits=0.03,%
        enlarge y limits=0.1,%
        nodes near coords, %
        point meta=explicit symbolic,%
        every axis plot/.style={very thick,mark=*,mark repeat=3,mark size=1.25pt},%
    }%
    \hfill%
    \begin{minipage}[t]{0.33\linewidth}%
        \begin{tikzpicture}%
            \begin{axis}[%
                title={Gems Environment},%
                legend to name=leg:plot1,%
                name=plot,%
                xlabel={Training Steps},%
                ylabel={Reward Recovery},%
                ymode=log,%
            ]%
                \makeplots\gemsdata%
            \end{axis}%
        \end{tikzpicture}%
    \end{minipage}%
    \hspace*{2ex}
    \begin{minipage}[t]{0.33\linewidth}%
        \begin{tikzpicture}%
            \begin{axis}[%
                title={Overcooked Environment},%
                name=plot,%
                legend to name=leg:plot2,%
                xlabel={Training Steps},%
                ymode=log,%
            ]%
            \makeplots\overcookeddata%

            \end{axis}%
        \end{tikzpicture}%
    \end{minipage}%
    \hspace*{-3ex}
    \begin{minipage}[t]{0.33\linewidth}%
        \begin{tikzpicture}%
        \begin{axis}[%
            title={Highway Environment},%
            name=plot,%
            xlabel={Training Steps},%
            legend to name=leg:plot3,%
            ymode=log,%
        ]%
            \makeplots\highwaydata% 
        \end{axis}%
        \end{tikzpicture}%
    \end{minipage}%
    \hfill%
    \vbox{\medskip\ref*{leg:plot}}%
	\caption{Reward recovery (MSE between true and predicted rewards) across training trajectories of the MAIRL algorithms.}
	\label{fig:reward_recovery}
\end{figure*}
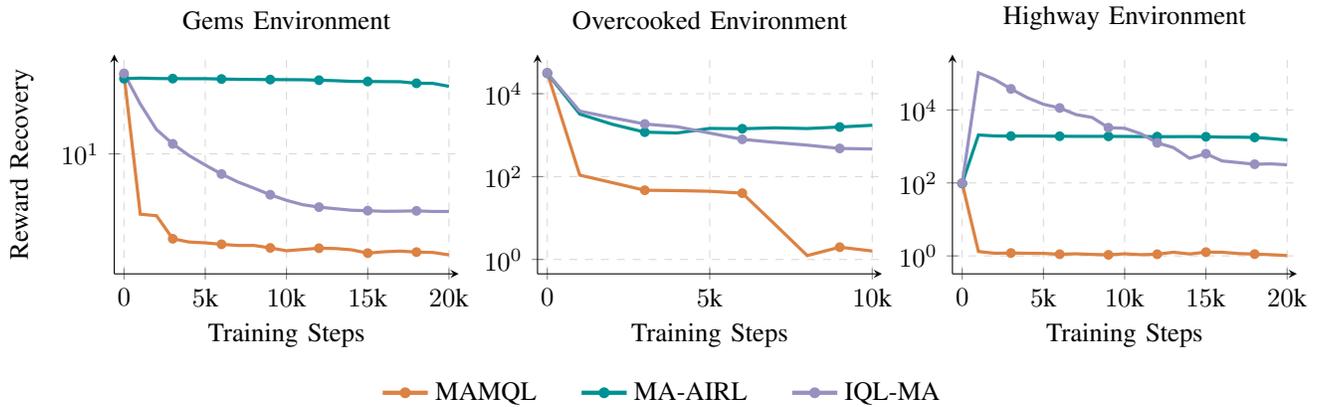

Furthermore, we compare \setup\ to IQ-Learn MA using recovered rewards, i.e., the reward functions learned by the algorithms.
\Cref{fig:reward_recovery} demonstrates the reward recovery of training trajectories of \setup\ and baselines where reward recovery is calculated from the MSE between true and predicted rewards.
\setup\ continues the observed rapid convergent behavior, improving reward alignment by 3-300$\times$ compared to baselines, where alignment performance is more pronounced on environments of increased complexity.

\pgfplotstableread{
step	0	1000	2000	3000	4000	5000	6000	7000	8000	9000	10000	11000	12000	13000	14000	15000	16000	17000	18000	19000	20000
MAMQL         57.19    1.23     1.06     0.88     0.72     0.68     0.67     0.66     0.66     0.65      0.64    0.65    0.63    0.61    0.61    0.60    0.59    0.58    0.56    0.54    0.52
IQL-MA        56.77    44.17    37.72    22.13    16.36    16.19    14.28    12.32    11.09    10.45     10.57   10.26   9.59    8.95    8.95    8.58    8.65    8.53    8.25    7.43    6.71
MA-AIRL       55.40    13.90    13.74    13.70    13.45    13.42    13.44    13.41    13.44    13.45     13.42   13.40   13.39   13.28   13.26   13.13   13.16   13.14   13.12   13.12   13.10
MAMQL_ERR     1.38     0.03     0.03     0.02     0.02     0.02     0.02     0.02     0.02     0.02      0.02    0.02    0.02    0.01    0.01    0.01    0.01    0.01    0.01    0.01    0.01
IQL-MA_ERR    1.37     1.06     0.91     0.53     0.39     0.39     0.34     0.30     0.27     0.25      0.26    0.25    0.23    0.22    0.22    0.21    0.21    0.21    0.20    0.18    0.16
MA-AIRL_ERR   1.33     0.33     0.33     0.33     0.32     0.32     0.32     0.32     0.32     0.32      0.32    0.32    0.32    0.32    0.32    0.32    0.32    0.32    0.32    0.32    0.32
}\gemsdata

\pgfplotstableread{
step          0        1000     2000     3000     4000     5000     6000     7000     8000     9000      10000
MAMQL         205.81   13.97    13.58    12.30    14.47    15.18    15.00    13.67    6.75     6.28      3.91
IQL-MA        208.81   209.72   230.67   299.51   343.85   517.80   641.75   789.55   978.77   1239.13   1408.97
MA-AIRL       206.05   249.46   253.25   255.31   255.84   257.16   258.87   257.89   258.81   259.04    259.35
MAMQL_ERR     0.18     0.01     0.01     0.01     0.01     0.01     0.01     0.01     0.01     0.01      0.00
IQL-MA_ERR    0.19     0.19     0.21     0.27     0.31     0.46     0.57     0.71     0.88     1.11      1.26
MA-AIRL_ERR   0.18     0.22     0.23     0.23     0.23     0.23     0.23     0.23     0.23     0.23      0.23
}\overcookeddata

\pgfplotstableread{
step          0        1000     2000     3000     4000     5000     6000     7000     8000     9000      10000   11000   12000   13000   14000   15000   16000   17000   18000   19000   20000
MAMQL         48.53    1.34     1.19     1.15     1.10     1.09     1.07     1.06     1.04     1.00      0.99    0.98    0.98    0.97    0.97    0.96    0.95    0.95    0.94    0.94    0.93
IQL-MA        47.25    2.08     2.02     1.98     1.91     1.90     1.86     1.81     1.80     1.73      1.71    1.63    1.61    1.57    1.55    1.51    1.41    1.43    1.43    1.36    1.31
MA-AIRL       48.26    32.82    25.83    22.88    22.82    22.81    22.77    22.14    21.83    21.19     21.06   21.00   20.66   20.53   20.43   20.38   20.07   19.96   19.88   19.40   19.12
MAMQL_ERR     1.01     0.03     0.02     0.02     0.02     0.02     0.02     0.02     0.02     0.02      0.02    0.02    0.02    0.02    0.02    0.02    0.02    0.02    0.02    0.02    0.02
IQL-MA_ERR    0.98     0.04     0.04     0.04     0.04     0.04     0.04     0.04     0.04     0.04      0.04    0.03    0.03    0.03    0.03    0.03    0.03    0.03    0.03    0.03    0.03
MA-AIRL_ERR   1.00     0.68     0.54     0.48     0.48     0.48     0.47     0.46     0.45     0.44      0.44    0.44    0.43    0.43    0.42    0.42    0.42    0.41    0.41    0.40    0.40
}\highwaydata

\pgfplotstabletranspose[colnames from=step,input colnames to=step]\gemsdata{\gemsdata}
\pgfplotstabletranspose[colnames from=step,input colnames to=step]\overcookeddata{\overcookeddata}
\pgfplotstabletranspose[colnames from=step,input colnames to=step]\highwaydata{\highwaydata}

\begin{figure*}[htb!]%
	\centering%
    \pgfplotsset{%
        xticklabel={%
            \pgfmathparse{\tick}%
            \pgfmathtruncatemacro\trunc{\pgfmathresult}%
            \pgfkeys{/pgf/fpu=true}%
            \ifnum\trunc<1000%
                \pgfmathprintnumber{\trunc}%
            \else%
                \pgfmathparse{\pgfmathresult/1000}%
                \ifnum\trunc<1000000\relax%
                    \pgfmathprintnumber{\pgfmathresult}k%
                \else%
                    \pgfmathparse{\pgfmathresult/1000}%
                    \pgfmathprintnumber{\pgfmathresult}M%
                \fi%
            \fi%
            \pgfkeys{/pgf/fpu=false}%
        },%
        grid=major,%
        grid style={gray!30,dashed},%
        width=1.05\linewidth,%
        height=4.5cm,%
        legend cell align=left,%
        axis lines=left,%
        scaled x ticks=false,%
        legend style={draw=none,at={(0.79,0.48)},anchor=center,/tikz/every even column/.append style={column sep=0.5cm},legend image post style={ultra thick}},%
        legend columns=-1,%
        enlarge x limits=0.03,%
        enlarge y limits=0.1,%
        nodes near coords, %
        point meta=explicit symbolic,%
        every axis plot/.style={very thick,mark=*,mark repeat=3,mark size=1.25pt},%
    }%
    \hfill%
    \begin{minipage}[t]{0.33\linewidth}%
        \begin{tikzpicture}%
            \begin{axis}[%
                title={Gems Environment},%
                legend to name=leg:plot1,%
                name=plot,%
                xlabel={Training Steps},%
                ylabel={Behavioral Error},%
                ymode=log,%
            ]%
                \makeplots\gemsdata%
            \end{axis}%
        \end{tikzpicture}%
    \end{minipage}%
    \hspace*{2ex}
    \begin{minipage}[t]{0.33\linewidth}%
        \begin{tikzpicture}%
            \begin{axis}[%
                title={Overcooked Environment},%
                name=plot,%
                legend to name=leg:plot2,%
                xlabel={Training Steps},%
                ymode=log,%
            ]%
            \makeplots\overcookeddata%

            \end{axis}%
        \end{tikzpicture}%
    \end{minipage}%
    \hspace*{-3ex}
    \begin{minipage}[t]{0.33\linewidth}%
        \begin{tikzpicture}%
        \begin{axis}[%
            title={Highway Environment},%
            name=plot,%
            xlabel={Training Steps},%
            legend to name=leg:plot3,%
            ymode=log,%
        ]%
            \makeplots\highwaydata% 
        \end{axis}%
        \end{tikzpicture}%
    \end{minipage}%
    \hfill%
    \vbox{\medskip\ref*{leg:plot}}%
	\caption{Behavioral error across trajectories while training the MAIRL algorithms.}
	\label{fig:behavioral_error}
\end{figure*}
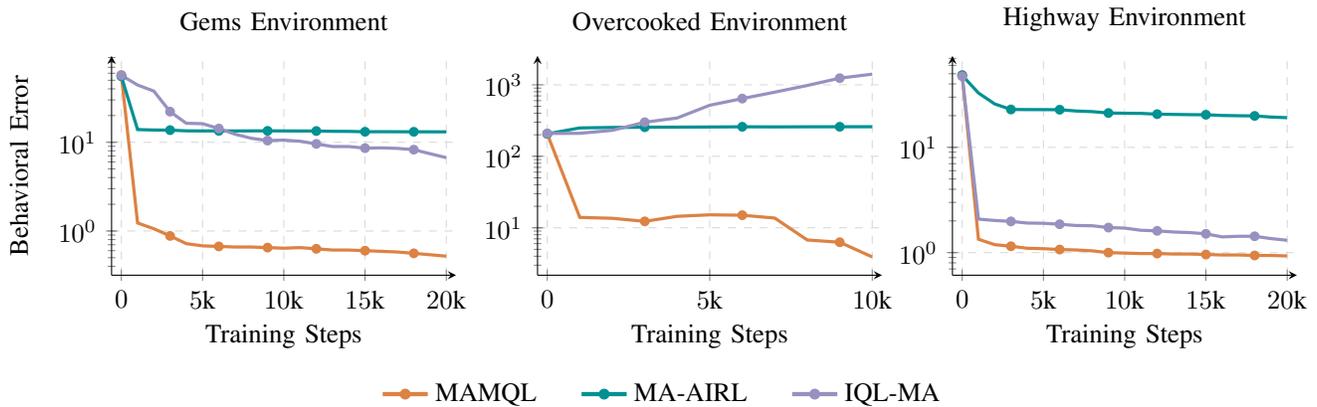

In addition to reward recovery, we also evaluate behavioral error, which measures the divergence between the actions taken by the learned policy and those taken by the expert policy. Figure \ref{fig:behavioral_error} illustrates that \setup\ consistently achieves lower behavioral error across all environments, outperforming the baselines significantly. Specifically, \setup\ demonstrates 20-66$\times$ lower behavioral error than the baselines, achieving higher fidelity in policy imitation.

We also observed that \setup\ was able to converge to the optimal policy with minimal hyperparameter modifications across environments: we only had to scale the buffer size to approximately 400$\times$ the horizon length of the environment. Furthermore, we observe \setup\ converges with a maximum buffer size of $16.6$k steps while other MAIRL algorithms in comparable environments, such as MAAC \cite{lowe2017multi} and MA-DAC \cite{xue2022multiagent}, were reported to necessitate buffer sizes approximately 3-75$\times$ larger.

\section{Conclusion}
We propose MAMQL, a novel multi-agent inverse reinforcement learning algorithm which addresses the complexities of learning rewards in multi-agent general-sum games. Our algorithm jointly learns reward functions and policies for each agent. Our results show the policies learned by MAMQL yield a significant increase in average reward and a decrease in training time across several environments compared to traditional MAIRL and IL algorithms. We demonstrate applications in autonomous driving and long-horizon tasks, though sim-to-real transfer mandates safety considerations and significant testing before deployment. One direction for future work is to explore expressive architectures further to exceed expert performance on higher-complexity tasks.

Another limitation of our work is experimental: we conducted our experiments using data from ``simulated'' experts. While our results are a proof of concept across different domains, humans may have biases in their decision making that cannot be captured by generalized Boltzmann policies \cite{laidlaw2021boltzmann,kwon2020humans}. Although these simplified computational models are common in inverse reinforcement learning, future work may focus on capturing humans' biases while learning policies or reward functions from their demonstrations, as it was done by Chan et al.~\citet{chan2021human} for single-agent systems.

\section*{Acknowledgments}
The authors would like to acknowledge Supervised Program for Alignment Research (SPAR) that helped form the team for this project.

\bibliographystyle{IEEEtran}
\balance\bibliography{references}

\end{document}